\documentclass{elsart3p}

\usepackage{graphicx}

\usepackage{amssymb}

\begin{document}

\begin{frontmatter}



\title{Uniqueness of Bohmian Mechanics, and Solutions From Probability Conservation}



\author[physics]{Timothy M. Coffey},
\ead{tcoffey@physics.utexas.edu}
\author[chemistry]{Robert E. Wyatt},
\author[physics]{Wm. C. Schieve}

\address[physics]{Department of Physics and Center for Complex Quantum Systems, 1 University Station C1600, University of Texas, Austin, TX 78712, USA}

\address[chemistry]{Department of Chemistry and Biochemistry and Institute for Theoretical Chemistry, 1 University Station A5300, University of Texas, Austin, TX 78712, USA}

\begin{abstract}
We show that one-dimensional Bohmian mechanics is unique, in that, the Bohm trajectories are the only solutions that conserve total left (or right) probability. In Brandt et al., {\it Phys. Lett. A}, {\bf 249} (1998) 265--270, they define {\it quantile motion}---unique trajectories are solved by assuming that the total probability on each side of the particle is conserved. They argue that the quantile trajectories are identical to the Bohm trajectories. Their argument, however, fails to notice the gauge freedom in the definition of the quantum probability current. Our paper sidesteps this under-determinedness of the probability current. The one-dimensional probability conservation can be used for higher dimensional problems if the wave function is separable. Several examples are given using total left probability conservation, most notably, the two-slit experiment.
\end{abstract}

\begin{keyword}
quantile motion \sep Bohm trajectories \sep two-slit experiment \sep separability

\PACS 03.65 \sep 03.65.Ta \sep 03.67.Lx
\end{keyword}
\end{frontmatter}

\section{Introduction}
\label{intro}
Bohm's formulation of quantum mechanics admits particle trajectories \cite{Bohm1952, Bohm1993, Holland1993}. Given a solution $\psi(x,t)=R(x,t)e^{i S(x,t)/\hbar}$ of Schr\"odinger's wave equation for real $R(x,t)$ and $S(x,t)$, Bohm defines the equation of motion for the particles as,
\begin{equation}
\label{secondlaw}
	m\ddot {\bf x} =-\nabla V - \nabla Q, 
\end{equation}
where the quantum potential is $Q=-\hbar^2 \nabla^2 R/(2m R)$. In order that an ensemble of particle trajectories satisfy the predictions of standard quantum mechanics, the particle velocities are further constrained to obey the so called {\it guidance law},
\begin{equation}
\label{guidancelaw}
	\dot {\bf x}=\frac{1}{m}\nabla S.
\end{equation}

The guidance law guarantees the equivalence of the particle ensemble density $\rho_{\rm p}$ and the quantum probability density $\rho_{\psi}$, since both densities evolve according to a continuity equation. From Schr\"odinger's equation the quantum probability density has,
\begin{equation}
\label{psicont}
	\frac{\partial \rho_\psi}{\partial t}
	+ \nabla\cdot
	\left(\rho_\psi\frac{\nabla S}{m}\right)
	=0,
\end{equation}
and as a result of particle conservation,
\begin{equation}
\label{particlecont}
	\frac{\partial \rho_{\rm p}}{\partial t}
	+ \nabla\cdot
	\left(\rho_{\rm p}\dot {\bf x}\right)=0.
\end{equation}

Therefore, if $\rho_{\rm p}(x,t=0)=\rho_\psi(x,t=0)$ then by the expressions above the two densities will remain identical. However, the definition of the guidance law [and in turn the equation of motion, Eq.~(1)] is not unique, since one can add a divergence-less gauge to the definition of the particle guidance law Eq.~(\ref{guidancelaw}), and the equivalence of the particle density $\rho_{\rm p}$ and the probability density $\rho_\psi$ is maintained as before. Hence, Bohm's formulation of quantum mechanics is under-determined \cite{Squires1996, Passon2005}. There has been some efforts to resolve this problem \cite{Deotto1998, Holland1999, Peruzzi1999, Stryve2004}. Either the proposal doesn't work, or if it does, it must add additional physical assumptions to Bohm's theory. In Brandt et al. \cite{Brandt1998, Brandt2001}, however, they give a non-physical reason for choosing Bohm's definition of the velocity field in one-dimension. They introduce the concept of {\it quantile motion}. A particle trajectory $x_P(t)$ is generated such that,
\begin{equation}
\label{CPF}
	P=\int_{-\infty}^{x_P(t)} \rho(x,t)\; dx = {\rm constant}.
\end{equation}
This is recognized also as the cumulative probability function (CPF) for the probability density $\rho(x,t)$. Additionally, one can think of this expression as representing the total left probability during the particle's trajectory. Note, since the total probability is conserved in time, one could have used the total right probability instead.

Their argument for the equivalence of the quantile trajectories and Bohm's trajectories rests upon the similarity between the two expressions for the velocity fields, 
\begin{equation}
	\dot x_P = \frac{j_{\rm p}}{\rho}
	\qquad{\rm and}\qquad
	\dot x_{\rm Bohm}= \frac{j_\psi}{\rho}.
\end{equation}
Where $j_{\rm p}$ is the particle current, and $j_\psi$ is the quantum probability current. They assume that the two currents are equal, hence, the quantile trajectories and Bohm's trajectories will be identical (if the initial conditions are the same). The two currents, however, can be related by a divergence-less gauge as shown above. Therefore, they fail to show that the quantile trajectories are in fact the Bohm trajectories because of the gauge freedom in the definition of the two currents.

Below we present a proof that sidesteps this under-determinedness of the currents by not introducing the current concept in the first place. Working directly from the quantile motion idea and Schr\"odinger's wave equation, we show that indeed the unique quantile trajectories for quantum mechanics are the Bohm trajectories. This gives a non-physical reason for selecting Bohm's definition of the velocity field (in one-dimension, at least), namely, the conservation of the total left (or right) probability.

Later, we show that the one-dimensional total left (or right) probability conservation can be used to generate Bohm trajectories for higher dimensional problems if the wave function is separable. Finally, we provide several comparisons of the quantile and Bohmian trajectories: harmonic oscillator, free particle, the two-slit experiment, and a two-dimensional infinite square well for a separable wave function. In all cases, the quantile trajectories are identical to the Bohm trajectories.

\section{Unique Quantile Trajectories in Quantum Mechanics are Bohmian}
\label{main}

In Brandt et al. \cite{Brandt1998, Brandt2001}, they show that one can construct trajectories by requiring that the total left (or right) probability is conserved along a particular trajectory. In their paper, they work with the total right probability,
\begin{equation}
	Q=\int_{x_Q(t)}^{+\infty} \rho(x,t)\; dx.
\end{equation}
But for the following we use the total left probability (noting that $P+Q=1$ for all time),
\begin{equation}
\label{totalleft}
	P=\int_{-\infty}^{x_P(t)} \rho(x,t)\; dx.
\end{equation}
This is also known as the cumulative probability function (CPF) for the probability density $\rho(x,t)$. The CPF is one-to-one and monotonically increasing. At each time, there is only one $x_p$ that satisfies Eq.~(\ref{totalleft}) for a constant value of $P$. Therefore, there is a {\it unique} trajectory such that the total left probability is conserved. Given that $P$ is constant, 
\begin{equation}
	\frac{dP}{dt}=\int_{-\infty}^{x_P(t)} 
		\frac{\partial \rho}{\partial t}\; dx
		+ \rho(x_P,t)\dot x_P =0.
\end{equation}
Where we assumed that the density is zero at the lower boundary. At once this can be solved for the velocity field that yields unique trajectories that conserve total left probability,
\begin{equation}
\label{xdot}
	\dot x_P = -\frac{1}{\rho(x_P,t)}\int_{-\infty}^{x_P(t)} 
		\frac{\partial \rho}{\partial t}\; dx.
\end{equation}

So far the discussion has been quite general and Eq.~(\ref{xdot}) is a definition for trajectories for any given density $\rho(x,t)$, {\it whether it be quantum or not}. To apply these quantile trajectories to quantum mechanics, we begin with Schr\"odinger's wave equation in one-dimension,
\begin{equation}
	i\hbar \frac{\partial \psi}{\partial t}
	= -\frac{\hbar^2}{2m}\frac{\partial^2 \psi}{\partial x^2}
	+ V(x)\psi,
\end{equation}
and the similar expression for the complex conjugate wave equation.
Writing the complex wave functions in polar form, $\psi(x,t)=R(x,t)e^{iS(x,t)/\hbar}$ for real $R$ and $S$, and following the typical derivation of the quantum continuity equation, we get that,
\begin{equation}
\label{nocurrent}
	-\frac{\partial \rho}{\partial t}
	=\frac{\partial \rho}{\partial x} \frac{\partial S}{m\partial x}
	+ \rho \frac{\partial^2 S}{m\partial x^2}.
\end{equation}
Notice, that we did not write the right hand side of this continuity equation as $\partial j/\partial x$ where $j=\rho \partial S/m\partial x$---this would have introduced the under-determinedness of the probability current since there is no unique anti-derivative. Inserting Eq.~(\ref{nocurrent}) into Eq.~(\ref{xdot}),
\begin{equation}
	\dot x_P = \frac{1}{\rho(x_P,t)}\int_{-\infty}^{x_P(t)} 
		\left(
		\frac{\partial \rho}{\partial x} \frac{\partial S}{m\partial x}
	+ \rho \frac{\partial^2 S}{m\partial x^2}
	\right)\; dx,
\end{equation}
and integrating by parts the second term of the integrand (again assuming that the density is zero at the lower boundary),
\begin{equation}
\label{bohmspeed}
	\dot x_P = \frac{\partial S}{m\partial x}.
\end{equation}
Which is the one-dimensional Bohm velocity field of Eq.~(\ref{guidancelaw}). We conclude, therefore, that the unique one-dimensional quantile trajectories are in fact the Bohm trajectories in quantum mechanics.

\section{Extension to Higher Dimensions}
\label{higher}

The extension of the quantile motion into higher dimensions is also discussed in Brandt et al. \cite{Brandt1998}. They show that instead of the total left (or right) probability being conserved in one-dimension, that in higher dimensions the probability is conserved inside a volume that is enclosed by a surface of Bohmian trajectories. This property, however, is not unique to Bohmian mechanics. Any velocity field $\dot{\bf x}$ will conserve the probability inside the volume in configuration space since \cite{Wyatt2005},
\begin{equation}
	\frac{d\rho}{dt}=-\rho\nabla\cdot \dot{\bf x}
	\quad{\rm and}\quad
	\frac{dJ}{dt}=+J\nabla\cdot \dot{\bf x},
\end{equation}
where $\rho$ is the probability density and $J$ is the Jacobian such that the volume changes as $dV(t)=J dV_0$. The probability inside this evolving volume is,
\begin{equation}
	P_{\rm in}=\int \rho \; dV(t) = \int \rho J \; dV_0. 
\end{equation}
Which implies that $dP_{\rm in}/dt = 0$. Therefore, any velocity field $\dot{\bf x}$ will conserve the total probability inside a volume that in enclosed by a surface of trajectories solved from $\dot{\bf x}$. This is in contrast to what was found in the one-dimensional case in \S\ref{main}, where there was a unique velocity field that satisfied the total left (or right) probability conservation.

However, the quantile motion concept can be used to generate trajectories in higher dimensions if one uses the marginal distribution for each coordinate analogously to the cumulative probability distribution (CPF) in one-dimension. Suppose the system can be described by $N$ Cartesian coordinates, then the corresponding definition of Eq.~(\ref{totalleft}) is,
\begin{equation}
\label{newtotalleft}
	P_i = \int_{-\infty}^{x_i(t)}\rho_i(x_i,t)\;dx_i
	\qquad{\rm for}\quad{i=1,2,\dots,N},
\end{equation}
where $\rho_i(x_i,t)$ is the marginal distribution for the $i$-th coordinate. Again, we assume that the particles are conserved so,
\begin{equation}
	\frac{\partial \rho}{\partial t}
	+\sum_{i=1}^N \frac{\partial}{\partial x_i}(\rho \dot x_i)
	=0.
\end{equation}
Partially, integrating this continuity equation we have that,
\begin{eqnarray}
	&&\int_{-\infty}^{+\infty}..\int_{-\infty}^{x_i(t)}..\int_{-\infty}^{+\infty}\frac{\partial \rho}{\partial t}\;dx_1\dots dx_N\nonumber
\\
	&&+\sum_{j=1}^N \int_{-\infty}^{+\infty}..\int_{-\infty}^{x_i(t)}..\int_{-\infty}^{+\infty}\frac{\partial (\rho \dot x_j)}{\partial x_j}\;dx_1\dots dx_N
	=0.
\end{eqnarray}
Interchanging the partial derivative with respect to time and performing the $\pm \infty$ integrations, and assuming that the density is zero at $\pm \infty$, only the $j=i$ integrals on the second term survive,
\begin{eqnarray}
	&&\int_{-\infty}^{x_i(t)}\frac{\partial \rho_i}{\partial t}\;dx_i\nonumber
	\\
	&&\qquad+\int_{-\infty}^{+\infty}..\int_{-\infty}^{+\infty}\rho \dot x_i \;dx_1\dots dx_{\neq i}\dots dx_N 
	=0.
\end{eqnarray}
Then with the expression above, and differentiating Eq.~(\ref{newtotalleft}) with respect to time,
\begin{equation}
	\frac{dP_i}{dt}= \rho_i \dot x_i
	-\int_{-\infty}^{+\infty}..\int_{-\infty}^{+\infty}\rho \dot x_i \;dx_1\dots dx_{\neq i}\dots dx_N.
\end{equation}
Hence, the $i$-th coordinate, in general, doesn't conserve total left probability of the marginal distribution since $\dot x_i$ could depend on the other coordinates. Suppose, however, that $\dot x_i=\dot x_i (x_i,t)$ (i.e. the motion along the $i$-th coordinate is independent), then $dP_i/dt=0$, and the total left probability is conserved for the marginal distribution $\rho_i$. 

In higher dimensional Bohmian problems the guidance law [Eq.~(\ref{guidancelaw})] for the $i$-th coordinate becomes,
\begin{equation}
	\dot x_i = \frac{1}{m}\frac{\partial S(x_1,x_2,\dots,x_N;t)}{\partial x_i}.
\end{equation}
If the wave function is separable, then,
\begin{equation}
	\psi=\psi_1(x_1,t)\psi_2(x_2,t)\cdots\psi_N(x_N,t).
\end{equation}
The probability density is also separable, $\rho=\rho_1(x_1,t)\rho_2(x_2,t) \dots\rho_N(x_N,t)$, and the phase becomes $S=S_1(x_1,t) + S_2(x_2,t)+\dots+S_N(x_N,t)$. Which implies that $\dot x_i = (1/m)\partial S_i(x_i,t)/\partial x_i$, or that the velocity field for the $i$-th coordinate is independent of the other coordinates, for all $i=1,2,\dots,N$. Hence, one can use the one-dimensional total left (or right) probability conservation described in \S\ref{main}, independently for each coordinate, to generate higher-dimensional Bohmian trajectories for a separable wave function.

\section{Examples}
\label{exam}

Below are several examples that compare the quantile trajectories to the Bohm trajectories. The first three examples are one-dimensional, while the last example is for a two-dimensional separable wave function. In each example, the wave function is non-stationary so that $\partial\rho/\partial t \neq 0$. From the wave function, the probability density is computed, $\rho(x,t)=|\psi(x,t)|^2$. The Bohmian trajectories can be numerically solved using Eq.~(\ref{bohmspeed}) or more conveniently from this alternative form for the velocity field \cite{Bohm1993, Holland1993},
\begin{equation}
	\dot x = \frac{\hbar}{2 m i}
	\left(\frac{\psi^\ast \partial \psi/\partial x
	-\psi \partial \psi^\ast /\partial x}
	{\rho}
	\right).
\end{equation}

The quantile trajectories are numerically computed by the method described in Appendix \ref{trapezoidmethod}. In the figures below, the quantile trajectory points ($+$) are shown against the Bohm trajectory. In all cases, quantile motion is able to reproduce the Bohm trajectories.

\subsection{Harmonic Oscillator}
\label{harm}

\begin{figure}
\begin{center}
\includegraphics[width=3in]{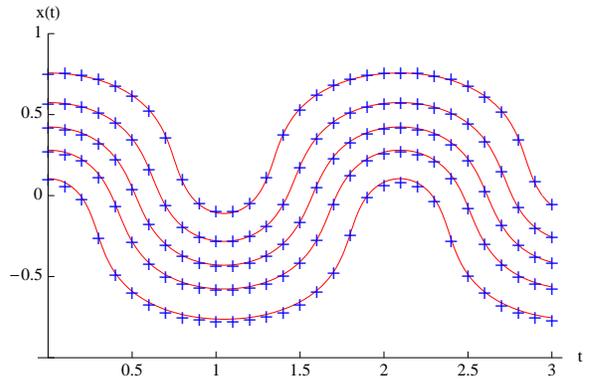}
\caption{(color online). Harmonic Oscillator with wave function as a superposition of the ground and first excited states. The quantile trajectory points ($+$) are shown superposed on the  Bohmian trajectories. The graph is in naturalized units.}
\label{harmfigure}
\end{center}
\end{figure}

In this example, the harmonic oscillator wave function was taken to be a superposition of the ground and first excited states,
\begin{eqnarray}
 	\psi(x,t)=\frac{1}{\sqrt{2}}
	&&\bigg(
	\sqrt{\frac{1}{a\sqrt{\pi}}}e^{-\frac{x^2}{2a^2}}e^{-i E_0 t/\hbar}\nonumber
	\\
	&&+\sqrt{\frac{1}{2 a \sqrt{\pi}}}
	e^{-\frac{x^2}{2a^2}}\frac{2 x}{a}e^{-i E_1 t/\hbar}
	\bigg),
\end{eqnarray}
where $a=\sqrt{\hbar/m\omega}$, and $E_j=\hbar\omega(j+1/2)$. Naturalized units were used so that $\hbar=1$, $\omega=3$, and $m=1$. The range of the time was $t=n\Delta t\in [0,3]$ for $n=1,2,3,\dots$, and the size of each time step was $\Delta t=0.1$. The cumulative probability function [the right hand side of Eq.~(\ref{totalleft})] was approximated by a series of trapezoids (see Appendix \ref{trapezoidmethod}) each with a width of $\Delta x=0.2$, and the position range was $x\in [-5,5]$ (an area where the density was essentially non-zero). In Fig.~(\ref{harmfigure}), the quantile trajectory points ($+$) are plotted superposed on top of the corresponding Bohm trajectories. We see that the quantile trajectories are, in fact, the Bohm trajectories. Smaller trapezoid widths and time steps will produce more accurate and continuous trajectories.

\subsection{Free Particle}
\label{free}

\begin{figure}
\begin{center}
\includegraphics[width=3in]{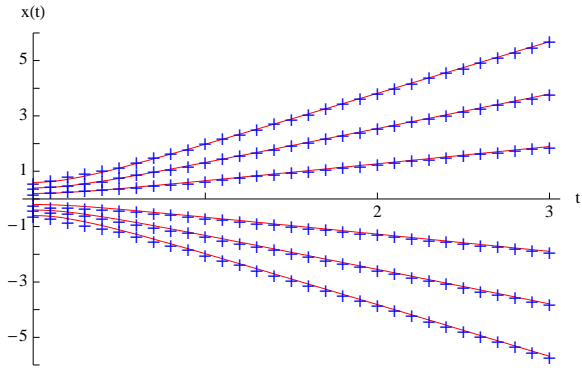}
\caption{(color online). Free particle with an initial wave function of a Gaussian centered around zero. The quantile trajectory points ($+$) are shown superposed on the  Bohmian trajectories. Even with only six trajectories shown the spreading of the wave packet is evident. The plot is using naturalized units.}
\label{freefigure}
\end{center}
\end{figure}

The wave function for the free particle (assumed to be Gaussian initially with width $a$) was taken to be,
\begin{equation}
	\psi(x,t)= \left(\frac{2 a}{\pi}\right)^{1/4}
	\frac{e^{-a x^2/[1+(2 i \hbar a t/m)]}}
	{\sqrt{1+(2 i\hbar a t/m)}}.
\end{equation}
Naturalized units were used so that $\hbar=1$, $m=1$, and $a=\pi/2$. Again, $t=n\Delta t\in [0,3]$ ($n=1,2,3,\dots$) with time steps of $\Delta t=0.1$. The trapezoid widths (see Appendix \ref{trapezoidmethod}) were $\Delta x=0.2$, while the range was $x\in [-5,5]$. In Fig.~(\ref{freefigure}), the quantile trajectory points ($+$) are shown against the Bohm trajectories. Notice that the ensemble of trajectories depict the familiar spreading of the wave function.

\subsection{Two-Slit Experiment}
\label{twoslits}

\begin{figure*}
\begin{center}
\includegraphics[width=6.5in]{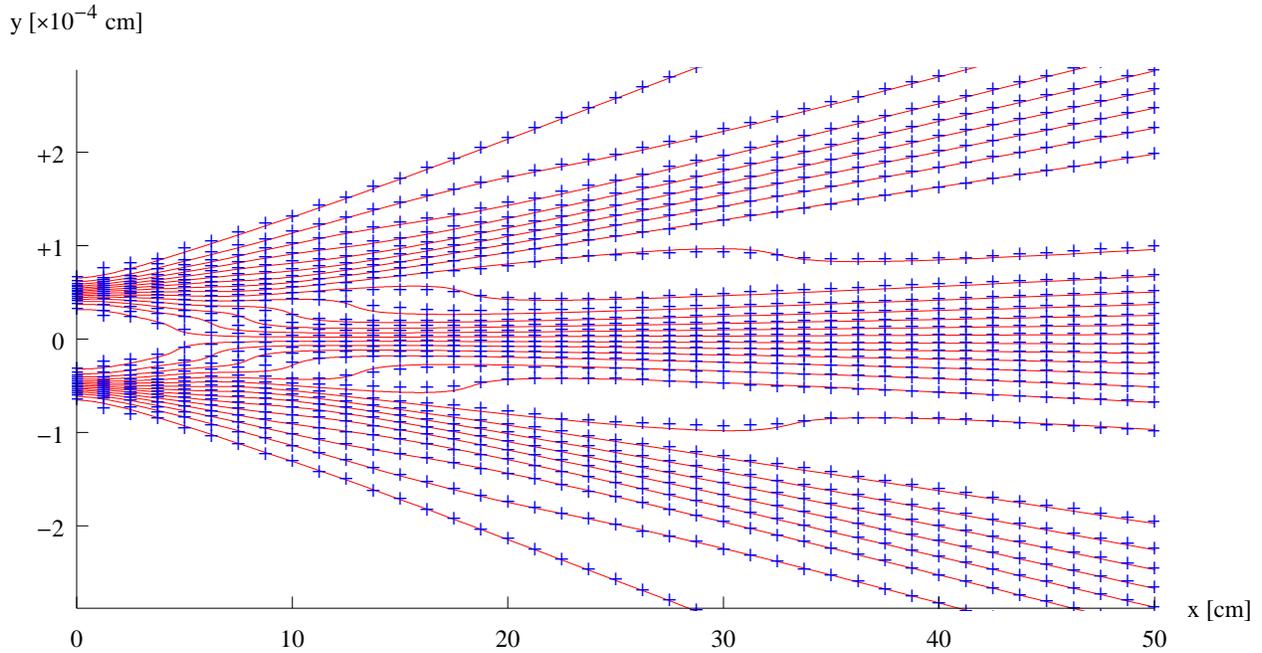}
\caption{(color online). Two-Slit Experiment as described in \S5.1.2 of P.R.~Holland, {\it The Quantum Theory of Motion: An Account of the de Broglie-Bohm Causal Interpretation of Quantum Mechanics}, (Cambridge University Press, New York, 1993). The quantile trajectory points ($+$) are shown superposed on the  Bohmian trajectories. The initial positions in each slit (left side of figure) are assumed to be Gaussian, and the ensemble of trajectories makes the familiar bands of bright and dark on the screen (right side of the figure).}
\label{twoslitsfigure}
\end{center}
\end{figure*}

This two-slit example is from \S5.1.2 in Holland \cite{Holland1993}. At first this problem seems to be two-dimensional. However, the motion along the coordinate from the slits to the screen [$x$ in Fig.~(\ref{twoslitsfigure})] is assumed uniform, thus the probability density is one-dimensional and is only a function of $y$ and $t$. To allow for easier calculation we rescaled the experimental numbers given in Holland, so that $\hbar =1$, $m=1$, and $t_{\rm max}=100$ (the time between the slits and the screen), and then rescaled the results back to the actual numbers. Using the trapezoid method as described in Appendix \ref{trapezoidmethod}, the time $t=n\Delta t\in[0,t_{\rm max}]$ ($n=1,2,3,\dots$) with a time step of $\Delta t=2.5$ ($1/40$-th the total time). At each time, the cumulative probability function (this time a function of $y$) was approximated by a series of trapezoids of width $\Delta y=3.24169$ ($1/80$-th the range of $y$). The range of $y$ was restricted to a width between $\pm 129.668$ where the probability density was essentially non-zero. In Fig.~(\ref{twoslitsfigure}), the quantile trajectory points ($+$) are plotted along with the Bohm trajectories. The ensemble of trajectories makes the familiar two-slit intensity pattern on the screen (located on the right hand side of the figure). The quantile trajectories match the Bohm trajectories quite well, even in those regions where the probability density is very close to zero (between the high intensity bands).

\subsection{Separable Wave Function in a Two-Dimensional Infinite Square Well}
\label{2dwell}

\begin{figure}
\begin{center}
\includegraphics[width=3in]{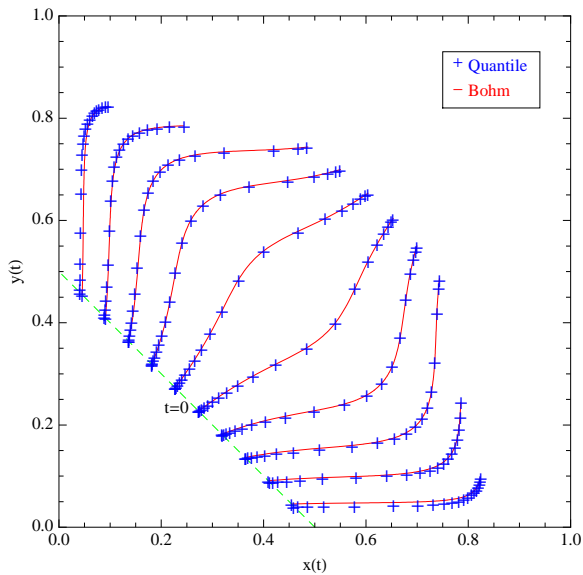}
\caption{(color online). Comparison of the quantile trajectories ($+$) and the Bohm trajectories for a separable wave function in the two-dimensional infinite square well of size $1\times 1$ in naturalized units. The initial position of each trajectory lies on the line from $(0.5,0)$ to $(0,0.5)$.}
\label{2dwellfigure}
\end{center}
\end{figure}

In Fig. (\ref{2dwellfigure}) is a comparison of the quantile trajectories and their Bohm counterpart for the two-dimensional infinite square well. The separable wave function was assumed to be $\psi(x,y,t)=\psi_x(x,t)\psi_y(y,t)$, where,
\begin{eqnarray}
	\psi_x(x,t)=\sqrt{\frac{1}{L}}
	&&\bigg(
	\sin\left(\frac{\pi x}{L}\right)e^{-i E_1 t/\hbar}\nonumber
	\\
	&&+ \sin\left(\frac{2 \pi x}{L}\right)e^{-i 4 E_1 t/\hbar}
	\bigg),
\end{eqnarray}
and a similar expression for $\psi_y(y,t)$. The energy $E_1=\pi^2 \hbar^2/2 m L^2$, and naturalized units were used so that $m=1$, $\hbar=1$, and the width of the well in each direction taken to be $L=1$. The time was $t=n\Delta t\in[0,1]$ for $n=1,2,3\dots$, and the size of each time step was $\Delta t=0.05$. The $(x(t),y(t))$ position of each particle was computed by approximating the cumulative probability function for each coordinate's marginal distribution by a series of trapezoids (see Appendix \ref{trapezoidmethod}) of width $\Delta x=\Delta y=L/30$. In Fig. (\ref{2dwellfigure}), the quantile trajectory points ($+$) are plotted superposed on top  of the corresponding Bohm trajectories. For the separable wave function, the quantile trajectories are again identical to the Bohm trajectories.

\section{Concluding Remarks}
\label{conclusion}

We showed that for one-dimensional probability densities one can define unique trajectories that conserve total left (or right) probability. These trajectories for one-dimensional quantum mechanics are the Bohm trajectories. What is interesting is that one can generate these one-dimensional Bohm trajectories without introducing any additional physical concepts (for example, the quantum potential). Hence, most of the example calculations showing Bohm trajectories that have appeared in the literature over the years, can be generated without any additional physical assumptions. In addition, using the quantile motion idea one can generate Bohm-like trajectories for {\it any} density, whether or not a quantum wave function is known or is exists.
In fact, the density could have been found experimentally, and the trajectories computed that depict its the evolution. The quantile motion idea can be used to generate higher dimensional trajectories for separable wave functions.

\section*{Acknowledgments}
R.E.W. thanks the Robert Welch Foundation for financial support.

\appendix

\section{Trapezoid Method for Generating Bohmian Trajectories 
From Probability Conservation}
\label{trapezoidmethod}

The method described below to generate one-dimensional Bohmian trajectories does {\it not} use either of Bohm's equations of motion, Eqs. (\ref{secondlaw}, \ref{guidancelaw}). Instead, the trajectories, $x_P(t)$, are computed using the quantile motion concept (or the conservation of the total left (right) probability) in Eq.~(\ref{totalleft}) for particular constant $P$ values. This expression, in general, can not be solved in closed form, and must be solved numerically.

\begin{figure}
\begin{center}
\includegraphics[width=3in]{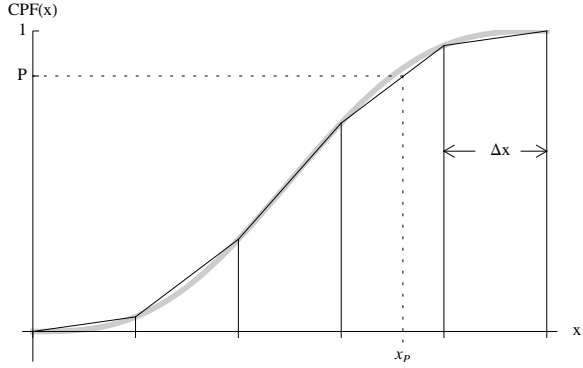}
\caption{At each time step the cumulative probability function (CPF) curve (in grey) is approximated by a series of trapezoids of width $\Delta x$. The position $x_P$,  corresponding to the constant quantile $P$ value, is found by solving the linear equation of the top line of the particular trapezoid. }
\label{CPFfigure}
\end{center}
\end{figure}

The trapezoid method approximates the cumulative probability function (CPF) with a series of equal width, $\Delta x$, trapezoids, see Fig.~(\ref{CPFfigure}). At each time $t=n\Delta t\ (n=1,2,3\dots)$ with time step $\Delta t$, the height of each side of every trapezoid is found by numerically integrating Eq.~(\ref{totalleft}). A particular quantile trajectory has a constant $P$ value between zero and one. From the $P$ value for a particular trajectory, the corresponding trapezoid is found. The $x_P$ value at each time is then calculated by solving the linear equation for the top segment of the corresponding trapezoid. As the number of trapezoids used for each time step increases, the better the approximation of the CPF curve, and hence the better the calculated position of the particle at the specific time step. And as the size of the time step is reduced the smoother the resulting trajectories. We found for the examples in \S\ref{exam} that $\Delta x/\Delta t\approx 3$ gave fair results.

\end{document}